\documentstyle{article}

\newcommand{\bea}{\begin{eqnarray}}
\newcommand{\eea}{\end{eqnarray}}
\newcommand{\beq}{\begin{equation}}
\newcommand{\eeq}{\end{equation}}
\newcommand{\nn}{\nonumber}

\def\k{{\vec k}}
\def\x{{\vec x}}
\def\/{\over}
\begin{document}

\parindent=1 em
\frenchspacing

\title{
Spontaneous excitation of an accelerated atom: \\ The contributions
of vacuum
fluctuations and radiation reaction}
\author{
\large J\"urgen Audretsch\thanks{e-mail:
audre@spock.physik.uni-konstanz.de}
 and Rainer M\"uller\thanks{e-mail:
rainer@spock.physik.uni-konstanz.de}
  \\
\normalsize \it Fakult\"at f\"ur Physik der Universit\"at Konstanz\\
\normalsize \it Postfach 5560 M 674, D-78434 Konstanz, Germany}
\date{\today}
\maketitle

\begin{abstract}
We consider an atom in interaction with a massless scalar quantum
field. We
discuss the structure of the rate of variation
of the atomic energy for an arbitrary stationary motion of the
atom through the quantum vacuum. Our main intention is
to identify and to analyze quantitatively the distinct contributions
of vacuum fluctuations and radiation reaction to the spontaneous
excitation
of a uniformly accelerated atom in its ground state. This gives an
understanding of the role of the different physical processes
underlying the
Unruh effect. The
atom's evolution into equilibrium and the Einstein coefficients for
spontaneous excitation and spontaneous emission are calculated.
\end{abstract}

\section{Introduction}

Spontaneous emission is one of the most prominent effects in the
interaction
of atoms with radiation. Two heuristic pictures have been put forward
to
explain why an atom in an excited state loses energy and radiates.
The first
one is inspired by classical electrodynamics: It is well known that,
classically, accelerated electrons in atoms radiate. The radiation
field
reacts back on the atom, causing a loss of atomic energy. This is
called
{\it radiation reaction}. The fact, however, that the rate of change
of an
atom's internal energy is always negative leads to the instability of
classical atoms with unacceptable consequences.

In a semiclassical theory, on the other hand, where quantum
mechanical
atoms interact with a classical radiation field, only stimulated
emission
and absorption are predicted; spontaneous emission is not present.
According to such a theory, excited atoms in the vacuum do not
radiate.
This has lead to the idea that spontaneous emission is connected with
the
quantum fluctuations of the radiation field. In particular,
spontaneous
emission has been interpreted as stimulated emission induced by {\it
vacuum fluctuations} \cite{Weisskopf35}. When making the argument
quantitative, however, the question arises: since stimulated emission
and
absorption have equal Einstein B coefficients, why do vacuum
fluctuations not
induce `spontaneous absorption' \cite{Milonni84}?

Quantum field theoretical investigations of the roles of vacuum
fluctuations
and radiation reaction in spontaneous emission have been carried out
since
1973 \cite{Ackerhalt73,Senitzky73,Milonni73,Ackerhalt74,Milonni75}.
A Heisenberg picture approach has always been used, since it allows
an easy
comparison of quantum mechanical and classical concepts. In these
studies,
the notion of vacuum fluctuations was connected with the free
solutions
of the Heisenberg equations for the quantum field, i. e. the field
that
is present even in the vacuum. Radiation reaction was incorporated
via the source field, which is the part of the field caused by the
presence of the
atom itself. Surprisingly, it turned out that seemingly the
contributions
of vacuum fluctuations and radiation reaction can to a large extent
be chosen
arbitrarily, depending on the ordering of commuting atom and field
variables.

Dalibard, Dupont-Roc, and Cohen-Tannoudji (DDC) argued in
\cite{Dalibard82} and \cite{Dalibard84} that there exists
a preferred operator ordering: Only if one choses a symmetric
ordering, the
distinct contributions of vacuum fluctuations and radiation
reaction to the rate of change of an atomic observable are separately
hermitean, and can possess an independent physical meaning. Using
this prescription, one finds that for an atom in an excited state,
vacuum
fluctuations and radiation reaction contribute equally to the rate of
change
of the atomic excitation energy. For an atom in the ground, state, on
the
other hand, the two contributions cancel precisely. There is a
balance
between vacuum fluctuations and radiation reaction that prevents
transitions
from the ground state and ensures its stability. In the same way, the
formalism of DDC can be used to study separately the effects of the
two physical mechanisms in a variety of situations. For example,
there
have been investigations of the radiative properties of atoms near a
conducting plane \cite{Meschede90,Meschede92} or of atomic level
shifts
in cavities \cite{Jhe91}. All these considerations refer to an atom
at rest.

In this paper, we want to study the modified effects of vacuum
fluctuations and
radiation reaction for an {\it accelerated} atom coupled to a quantum
field
in free space.
It is well known that a uniformly accelerated atom in its ground
state is
spontaneously excited even in the vacuum \cite{Unruh76} (cf. also
\cite{Davies75}). This process, which is called the
{\it Unruh effect}, is connected with the emission of a particle
\cite{Unruh84,Audretsch93b}. The intimate relation between
spontaneous
emission and Unruh effect has been noted previously
\cite{Audretsch93a}.

To understand the physical mechanisms responsible for the spontaneous
excitation of an accelerated atom, it
appears promising to apply the methods developed in the theory
of spontaneous emission for inertial atoms to the case of accelerated
atoms.
We will identify the contributions of vacuum fluctuations and
radiation
reaction to the rate of change of the atomic Hamiltonian in the
accelerated
case. As we will see, the effect of vacuum fluctuations on the atom
is
changed by the acceleration whereas the contribution of radiation
reaction
is completely unaltered. This leads to the following picture: For an
atom
in an excited state, there will be spontaneous emission with a
modified
transition rate. For an atom in the ground state, however, the
balance
between vacuum fluctuations and radiation reaction is no longer
perfect.
Due to the modified vacuum fluctuation contribution, transitions to
an excited state become possible even in the vacuum: this is the
Unruh effect. In working out the respective details, we substantiate
in a
quantitative way the intuitive picture described by Sciama, Candelas,
and
Deutsch \cite{Sciama81} generalizing the discussion in Ref.
\cite{Fain69}.
We mention that as far as the Unruh effect is concerned, an
alternative
discussion based on a classical notion of vacuum fluctuations has
been given in \cite{Boyer80}.

The paper is organized as follows: In Sec. 2, we introduce the model
of an
atom coupled to the radiation field. For simplicity, we chose a
two-level atom and a scalar quantum field. In Sec. 3, the Heisenberg
equations of motion are derived and formally solved. In Sec. 4, we
identify
the contributions of vacuum fluctuations and radiation reaction to
the
rate of change of an arbitrary atomic observable. We generalize the
formalism of DDC \cite{Dalibard84} to an atom in arbitrary stationary
motion.
The
special case of the evolution of the atom's excitation energy is
considered.
It is applied in Sec. 5 to the spontaneous emission from
an inertial atom in vacuum and in a heat bath. In Sec. 6, we treat a
uniformly accelerated atom and
discuss the physical reasons for its spontaneous excitation. Finally,
in
Sec. 7,
we derive the evolution of the atom's energy and the Einstein
coefficients
for spontaneous emission and Unruh effect.

\section{Interaction of a two-level atom and a scalar quantum field}

We want to study the interaction of a two-level atom and a real
scalar massless quantum field as a simplified model of quantum
electrodynamics.
$x \leftrightarrow (t,\x)$ are the Minkowski coordinates refering to
an
inertial reference frame.
We consider an atom on a stationary trajectory $x ( \tau) = (t(\tau),
\x(\tau))$, where
$\tau$ denotes its proper time. Throughout the paper, the time
evolution of the coupled system will be described with respect to
$\tau$. The stationary trajectory guarantees that the undisturbed
atom has
stationary states which
are called $| - \rangle$ and $|+ \rangle$, with energies $ - {1\/2}
\omega_0$ and $+ {1\/2} \omega_0$ and a level spacing $\omega_0$.
The Hamiltonian which governs the time evolution of the atom
with respect ot $\tau$ can then be written in Dicke's \cite{Dicke54}
notation
\beq H_A (\tau) =\omega_0 R_3 (\tau)  \label{eq1} \eeq
where $R_3 = {1\/2} |+ \rangle \langle + | - {1\/2}| - \rangle
\langle - |$
and $\hbar=c=1$.

The free Hamiltonian of the quantum field which generates the time
evolution
with regard to $t$ is given by
\beq {\tilde H}_F (t)  = \int d^3 k \, \omega_\k\, a^\dagger_\k a_\k.
\label{eq2}\eeq
$a^\dagger_\k$, $a_\k$ are the creation and annihilation operators
for a `photon' with momentum $\k$. ${\tilde H}_F (t)$
governs the time evolution of the field in $t$, the inertial time
of the laboratory system. We change to the new time variable $\tau$.
Heisenberg's equations of motion show that the Hamiltonian with
respect to
$\tau$ is given by
\beq H_F (\tau) = \int d^3 k\, \omega_\k \,a^\dagger_\k a_\k
	{dt\/d \tau}. \label{eq3}\eeq
The decomposition of the field operator in terms of creation and
annihilation operators reads
\beq \phi ( t, \x) =  \int d^3 k\, g_{\k} \left( a_\k (t) e^{i \k \x}
+
a^\dagger_\k (t) e^{-i \k \x} \right) \label{eq4}\eeq
where $g_\k = (2 \omega_\k (2 \pi)^3)^{-{1\/2}}$.

We couple the atom and the field by the scalar counterpart of the
electric dipole interaction
\beq H_I (\tau) = \mu R_2 (\tau) \phi ( x(\tau)). \label{eq5}\eeq
The coupling is effective only on the trajectory $x(\tau)$ of the
atom. $\mu$ is a coupling constant which we assume to be small.
$R_2$ is a matrix that connects only different atomic states:
$R_2 = {1\/2} i ( R_- - R_+)$, where $R_+ = |+ \rangle \langle - |$
and $R_- = |- \rangle \langle +|$ are the atomic raising and lowering
operators. The operators $R_3$ and $R_\pm$ obey angular momentum
algebra: $[R_3, R_\pm]= \pm R_\pm$, $[R_+, R_-] = 2 R_3$.

The Heisenberg equations of motion for the dynamical variables of
the atom and the field can be derived from the Hamiltonian $H= H_A
+ H_F +H_I$:
\beq {d \/ d \tau} R_\pm (\tau) = \pm i \omega_0 R_\pm (\tau) + i \mu
	\phi (x(\tau))[ R_2 (\tau), R_\pm (\tau) ], \label{eq6}\eeq
\beq {d\/ d \tau} R_3 (\tau) = i \mu \phi (x(\tau)) [R_2 (\tau),
	R_3 (\tau)], \label{eq7}\eeq
\beq {d\/dt} a_\k (t(\tau)) = - i \omega_\k a_\k (t(\tau)) + i \mu
R_2 (\tau)
	[ \phi (x(\tau)), a_\k (t(\tau))] {d \tau \/ dt}.
\label{eq8}\eeq
In Eq. (\ref{eq8}), $t$ must be considered as a function of $\tau$.
We
prefer to leave the commutators occurring in (\ref{eq6}) -
(\ref{eq8})
unevaluated because this will simplify the physical picture in the
later
sections.

The solutions of the equations of motion can be split into two parts:
(1) The {\it free part} which is present even in the absence of the
coupling,
(2) the {\it source part} which is caused by the interaction between
atom
an field and contains the coupling constant $\mu$:
\bea R_\pm (\tau) &=& R^f_\pm (\tau) + R^s_\pm (\tau), \nonumber \\
	R_3(\tau) &=& R^f_3 (\tau) + R_3^s (\tau), \nonumber \\
	a_\k (t(\tau)) &=& a_\k^f (t(\tau)) + a_\k^s
(t(\tau)).\nonumber\eea
Formal integration of the equations of motion yields
\beq  R^f_\pm (\tau) = R^f_\pm (\tau_0) e^{\pm i \omega_0 (\tau -
\tau_0)},
	\qquad  R^s_\pm (\tau) = i \mu \int_{\tau_0}^\tau d \tau' \,
\phi^f
	(x(\tau'))[ R_2^f (\tau'),R_\pm^f ( \tau) ], \label{eq9}\eeq
\beq R_3^f (\tau)= R_3^f (\tau_0), \qquad
	R_3^s (\tau) = i \mu \int_{\tau_0}^\tau d \tau' \, \phi^f
(x(\tau'))
	[R_2^f (\tau'), R_3^f (\tau) ], \label{eq10}\eeq
\beq a_\k^f (t(\tau)) = a_\k^f (t(\tau_0)) e^{ i \omega_\k (t(\tau)
	-t(\tau_0))}, \qquad
	a_\k^s (t(\tau)) = i \mu \int_{\tau_0}^\tau d \tau' \, R_2^f
(\tau')
	[ \phi^f (x(\tau')),a_\k^f (t(\tau)) ]. \label{eq11}\eeq
In the source parts of the solutions, all operators on the right hand
side
have been replaced by their free parts, which is correct to first
order
in $\mu$. From (\ref{eq11}), we can construct the free and source
part of the quantum field $\phi$:
\bea \phi^f (t(\tau), \x (\tau)) &=& \int d^3 k\, g_{\k} \left( a_\k
(0)
	e^{i \k \x(\tau)-i \omega_\k t(\tau)} + a^\dagger_\k (0)
	e^{-i \k \x(\tau) + i \omega_\k t(\tau)} \right), \nonumber\\
\phi^s (t(\tau), \x (\tau)) &=& i \mu \int_{\tau_0}^\tau d \tau' \,
	R_2^f (\tau') [\phi^f (x(\tau')),\phi^f
(x(\tau))].\label{eq12}\eea

\section{Vacuum fluctuations and radiation reaction}
\label{sec:vf+rr}

We assume
that the initial state of the field is the vacuum $|0 \rangle$, while
the atom is prepared in the state $|a \rangle$, which may be
$|+\rangle$
or $|-\rangle$. In principle, the time evolution of the mean value of
any atomic observable $G$ could
be calculated as the solution of a coupled set of Heisenberg
equations
analogous to to (\ref{eq6}) - (\ref{eq8}). Our aim is, however, to
identify and separate on the basis of (\ref{eq12}) in the rate of
change of
$G(\tau)$ the contributions that are caused by two distinct physical
mechanisms:
(1) The change in $G$ produced by the fluctuations of the quantum
field
which are present even in the vacuum. This part is related to the
{\it free} part of the field and is called the contribution of the
{\it vacuum fluctuations} to $dG\/d \tau$,
(2) the change in $G$ due to the interaction with that part of the
field
which is caused by the atom itself. This is the contribution of
{\it radiation reaction} to $dG\/d \tau$ and is connected with the
source part of the field.

The task of identifying the contributions of vacuum fluctuations and
radiation reaction in the dynamics of a small system coupled to a
reservoir has been considered by Dalibard, Dupont-Roc, and
Cohen-Tannoudji
(DDC) \cite{Dalibard82,Dalibard84}. We will apply their formalism to
the
problem of a two-level atom coupled to the radiation field and
generalize
it to arbitrary stationary trajectories $x(\tau)$ of the atom.

The Heisenberg equations of motion for an arbitrary atomic observable
$G(\tau)$ is given by
\beq {d\/d \tau} G(\tau) = i [H_A (\tau), G(\tau)] + i [ H_I (\tau),
	G (\tau)]. \label{eq13}\eeq
We are interested only in the part of $dG\/d \tau$ due to the
interaction
with the field:
\beq \left( {d G(\tau) \/ d \tau} \right)_{\hbox{coupling}} = i \mu
\phi (x(\tau))
	[R_2 (\tau), G (\tau)]. \label{eq14}\eeq
In order to identify the contributions of vacuum fluctuations and
radiation
reaction, we must investigate in (\ref{eq14}) the effects of $\phi^f$
and
$\phi^s$ separately.

At this point, however, an operator ordering problem arises. The
feature
that all atomic observables commute with $\phi$ is preserved in time
because of the unitary evolution. This is not true for $\phi^f$ and
$\phi^s$ separately. The reason for this is that the source part of
$\phi$
picks up contributions of atomic observables during its time
evolution
and vice versa (cf. (\ref{eq9}) - (\ref{eq11})). Because (\ref{eq14})
contains products of atomic and field operators, we must therefore
choose an operator ordering in (\ref{eq14}) when discussing the
effects
of $\phi^f$ and $\phi^s$ separately:
\beq
\left( {d G(\tau) \/ d \tau} \right)_{\hbox{coupling}} = i \mu \Bigl(
	\lambda\phi (x(\tau)) [R_2 (\tau), G (\tau)] +
(1-\lambda)[R_2 (\tau),
	G (\tau)]  \phi (x(\tau)) \Bigr), \label{eq14a}\eeq
with an arbitrary real $\lambda$. Different operator orderings will
lead
to the same final results for physical quantities on the l. h. s. of
(\ref{eq14a}), but will yield
different interpretations concerning the roles played by vacuum
fluctuations
and radiation reaction
\cite{Ackerhalt73,Senitzky73,Milonni73,Ackerhalt74,%
Milonni75}. However, DDC noticed that there exists a preferred
ordering
prescription: They showed that only if a symmetric ordering
($\lambda={1\/ 2}$)
of atomic
and field variables is adopted, $\left( {dG\/d \tau} \right)_{vf}$
and
$\left( {dG\/d \tau} \right)_{rr}$ are both hermitean. They argued
that
only under this condition, the effects of vacuum fluctuations and
radiation reaction can possess an independent physical meaning.

Adopting the symmetric ordering prescription in (\ref{eq14}), we can
identify the contribution of the {\it vacuum fluctuations} to $dG\/ d
\tau$,
\beq \left( {d G(\tau) \/ d \tau} \right)_{vf} = {1\/2}i \mu \Bigl(
\phi^f
	(x(\tau)) [R_2 (\tau), G (\tau)] + [R_2 (\tau), G (\tau)]
	\phi^f (x(\tau)) \Bigr), \label{eq15}\eeq
which goes back to $\phi^f$, and the contribution of {\it radiation
reaction},
\beq \left( {d G(\tau) \/ d \tau} \right)_{rr} = {1\/2}i \mu \Bigl(
\phi^s
	(x(\tau)) [R_2 (\tau), G (\tau)] + [R_2 (\tau), G (\tau)]
	\phi^s (x(\tau)) \Bigr) \label{eq16}\eeq
which goes back to the source part $\phi^s$ of the field.

We note that if the initial state of the field is not the vacuum but
some
large reservoir of $\phi$-particles (photons), the expression
(\ref{eq15})
represents the reservoir fluctuations $\left( {dG\/d \tau}
\right)_{rf}$.
The vacuum can be regarded as a particular reservoir.

\section{Rate of variation of the atomic energy in vacuum for
arbitrary
stationary motion}
\label{sec:rate}

We are now prepared to identify the contributions of vacuum
fluctuations and
radiation reaction in the evolution of the atom's excitation energy,
which is given by the expectation value of $H_A = \omega_0
R_3(\tau)$.
The free part of the atomic Hamiltonian is constant in time so that
the
rate of change of $H_A$ consists only of the two contributions
obtained
from (\ref{eq15}) and (\ref{eq16}):
\bea \left({d H_A (\tau) \/ d\tau} \right)_{vf} &=&
	{1\/2} i \omega_0 \mu \Bigl( \phi^f (x(\tau)) [R_2 (\tau),
R_3 (\tau)] +
	[R_2 (\tau), R_3 (\tau)] \phi^f (x(\tau)) \Bigr),
\label{eq17}\\
\left( {d H_A (\tau) \/ d\tau} \right)_{rr} &=& {1\/2} i \omega_0 \mu
\left(
	\phi^s (x(\tau)) [R_2 (\tau), R_3 (\tau)] + [R_2 (\tau), R_3
	(\tau)] \phi^s (x(\tau)) \right). \label{eq18}\eea
We can separate $R_2 (\tau)$ and $R_3(\tau)$ into their free part
(zeroth
order in $\mu$) and source part (first order in $\mu$), cf.
(\ref{eq9})
and (\ref{eq10}). In a perturbative treatment, we take into account
only
terms up to order $\mu^2$. We will also express all terms on the
right hand
side of (\ref{eq17}) and (\ref{eq18}) in terms of the free parts
$R_2^f$, $R_3^f$, and $\phi^f$.This will allow us to describe the
atom's
evolution with respect to simple statistical functions of the atom
and the
field. Therefore we use for the corresponding source parts the
expressions
(\ref{eq9}) - (\ref{eq11}) from the solutions of the Heisenberg
equations
above. We obtain up to order $\mu^2$
\bea \left( {d H_A (\tau) \/ d\tau} \right)_{vf}  &= &
	{1\/2} i \omega_0 \mu \Bigl( \phi^f (x(\tau)) [ R_2^f (\tau),
R_3^f (\tau)]
	+ [ R_2^f (\tau), R_3^f (\tau)] \phi^f (x(\tau)) \Bigr)
\label{eq19}\\
&&\qquad - {1\/2} \omega_0 \mu^2 \int_{\tau_0}^\tau d \tau'\left\{
\phi^f
	(x(\tau)), \phi^f (x(\tau'))\right\} \left[ R_2^f (\tau'), [
R_2^f (\tau),
	R_3^f (\tau)] \right], \nonumber\\
\left( {d H_A (\tau) \/ d\tau} \right)_{rr} &=&
	-{1\/2} \omega_0 \mu^2 \int_{\tau_0}^\tau d \tau'\, [\phi^f
(x(\tau)),
	\phi^f (x(\tau'))] \label{eq20}\\
&&\qquad \times \left( R_2^f (\tau') [R_2^f (\tau), R_3^f (\tau)]
	+ [R_2^f (\tau), R_3^f (\tau)] R_2^f (\tau') \right),
	\nonumber\eea
where curly brackets denote the anticommutator and we have used the
fact that free atom and field variables commute.

We are interested only in atomic observables. Consequently, we
perform an
averaging over the field degrees of freedom by taking the vacuum
expectation
value of (\ref{eq19}) and (\ref{eq20}). The right hand sides of
(\ref{eq19})
and (\ref{eq20}) contain only free operators so that only $\phi^f$
is affected. Since $\langle 0 | \phi^f | 0 \rangle =0$, the first
line
of (\ref{eq19}) does not contribute and we obtain
\bea \langle 0| {d H_A (\tau) \/ d\tau} |0\rangle_{vf} &=&
	- \omega_0 \mu^2 \int_{\tau_0}^\tau d \tau'\,
C^F(x(\tau),x(\tau'))
	\left[ R_2^f (\tau'), [R_2^f (\tau),R_3^f (\tau)]\right],
\label{eq21}\\
\langle 0| {d H_A (\tau) \/ d\tau} |0\rangle_{rr} &=&
	\omega_0 \mu^2 \int_{\tau_0}^\tau d \tau'\,
\chi^F(x(\tau),x(\tau'))
	\left\{ R_2^f (\tau'),[R_2^f (\tau),R_3^f (\tau)]\right\}.
	\label{eq22}\eea
The statistical functions $C^F$ and $\chi^F$ of the field are well
known
from linear response theory \cite{Martin67}. The {\it symmetric
correlation
function} of the field is given by
\beq C^{F}(x(\tau),x(\tau')) = {1\/2} \langle 0| \{ \phi^f (x(\tau)),
\phi^f
	(x(\tau')) \} | 0 \rangle, \label{eq23} \eeq
It is sometimes also called Hadamard's elementary function and
describes
the fluctuations of the quantum field in the vacuum state.
The {\it linear susceptibility} of the field is defined as
\beq \chi^F(x(\tau),x(\tau')) = {1\/2} \langle 0| [ \phi^f (x(\tau)),
\phi^f
	(x(\tau'))] | 0 \rangle \label{eq24}\eeq
and is also known as Pauli-Jordan of Schwinger function. According to
(\ref{eq22}), it describes the linear response of the averaged rate
${d H_A (\tau) \/ d\tau}$ on fluctuations of the atom.
Note that the statistical functions of the field have to be evaluated
along the trajectory of the atom.

We do not intend to deal with the operator dynamics of Eqs.
(\ref{eq21})
and (\ref{eq22}). Instead we are interested in the evolution of
expectation
values of atomic observables. Accordingly, we take the expectation
value
of Eqs. (\ref{eq21}) and (\ref{eq22}) in the atom's state $|a
\rangle$.
We can replace with (\ref{eq7}) to order $\mu^2$ the commutator
$\omega_0
[R_2^f (\tau), R_3^f (\tau)]$ with $i{d\/d \tau} R_2^f (\tau)$ and
obtain
as a central result
\bea \left\langle {d H_A (\tau) \/ d\tau} \right\rangle_{vf} &=&
	2 i \mu^2 \int_{\tau_0}^\tau d \tau' \, C^F(x(\tau),x(\tau'))
	{d\/ d \tau} \chi^A(\tau,\tau'), \label{eq25}\\
\left\langle {d H_A (\tau) \/ d\tau} \right\rangle_{rr} &=& 2 i \mu^2
	\int_{\tau_0}^\tau d \tau' \, \chi^F(x(\tau),x(\tau')) {d\/ d
\tau}
	C^A(\tau,\tau'). \label{eq26}\eea
where $\langle \dots \rangle = \langle 0, a| \dots |0,a \rangle$.
The rate of change of the atom's excitation energy is expressed
entirely
in terms of statistical functions. The statistical functions of the
atom are defined analogous to (\ref{eq23}) and (\ref{eq24}):
\bea C^{A}(\tau,\tau') &=& {1\/2} \langle a| \{ R_2^f (\tau), R_2^f
(\tau')\}
	| a \rangle \label{eq27} \\
\chi^A(\tau,\tau') &=& {1\/2} \langle a| [ R_2^f (\tau), R_2^f
(\tau')]
	| a \rangle. \label{eq28}\eea
$C^A$ is called the {\it symmetric correlation function} of the atom
in the state $|a \rangle$, $\chi^A$
its {\it linear susceptibility}. It will be important that $C^A$ and
$\chi^A$
do not depend on the trajectory of the atom but characterize only the
atom
itself.

The physical picture implied by Eqs. (\ref{eq25}) and (\ref{eq26})
can
be expressed in the following way \cite{Dalibard82}: (1) The field
fluctuates and acts on the atom, which is polarized (contribution of
the vacuum fluctuations), (2) The atom fluctuates and perturbs the
field,
which in turn reacts back on the atom (contribution of radiation
reaction).
The intimate relation between dissipative and fluctuative processes
provides an example for the fluctuation-dissipation theorem
\cite{Callen51,Milonni84}. Because of the symmetric operator ordering
chosen in Sec. \ref{sec:vf+rr}, only the commutator appears in
$\chi^F$
of (\ref{eq23}). Accordingly, the radiation reaction contribution
(\ref{eq26})
does not depend on the state of the field. This is plausible, since
radiation
reaction is connected only with the part of the field radiated by the
atom
and justifies again the choice of the symmetric ordering.

Finally, we want to give the explicit forms of the statistical
functions
of the atom and the field, which will be useful in the following
sections.
We obtain for the statistical functions of the atom:
\bea C^{A}(\tau,\tau') & =& {1\/2} \sum_b |\langle a | R_2^f (0) | b
	\rangle |^2 \left( e^{i \omega_{ab}(\tau - \tau')} + e^{-i
\omega_{ab}
	(\tau - \tau')} \right), \label{eq29}\\
\chi^A(\tau,\tau') & =& {1\/2} \sum_b |\langle a | R_2^f (0) | b
\rangle |^2
	\left(e^{i \omega_{ab}(\tau - \tau')} - e^{-i
\omega_{ab}(\tau - \tau')}
	\right), \label{eq30}\eea
where $\omega_{ab}= \omega_a-\omega_b$ and the sum extends over a
complete
set of atomic states.

The statistical functions of the field are well known from special
relativistic
quantum field theory \cite{Heitler54} with reference to Minkowski
coordinates
$t$ and $\x$. They can be written as functions of $\tau$
\bea C^{F}(x(\tau),x(\tau')) & =& {1\/8 \pi^2} {1\/| \x |} \left(
	{{\cal P}\/ \Delta t + |\Delta \x|} - {{\cal P}\/ \Delta
t-|\Delta \x|}
	\right) \nonumber\\
	&=& - {1\/ 8\pi^2} \left( {1 \/(\Delta t+ i \epsilon)^2 -
|\Delta \x|^2} +
	{1 \/(\Delta t- i \epsilon)^2 - |\Delta \x|^2}\right),
\label{eq31}\\
\chi^F(x(\tau),x(\tau')) & =& { i\/ 8 \pi}{1\/ |\Delta \x|} \left(
\delta(
	\Delta t + \Delta|\x|)- \delta(\Delta t - |\Delta \x|)
\right), \label{eq32}\eea
where $\Delta t=t(\tau) - t(\tau')$, $\Delta \x = \x (\tau) - \x
(\tau')$
and $\cal P$ denotes the principal value.

\section{Spontaneous emission from a uniformly moving atom}
\label{sec:inertial}

In this section, we apply the previously developed formalism to the
well
known problem of the spontaneous emission from an inertially moving
atom.
This will provide a basis for the discussion of the role of vacuum
fluctuations
and radiation reaction in the more general case of accelerated atoms
in the
following section. We consider an atom with constant velocity $\vec
v$
on the trajectory
\beq  t(\tau)  = \gamma \tau, \qquad
	\x (\tau)  = \x_0 + {\vec v} \gamma \tau, \label{eq33}\eeq
where $\gamma = (1-v^2)^{-{1\/2}}$. The statistical functions of the
field are easily calculated from their general forms (\ref{eq31}) and
(\ref{eq32}):
\bea C^F(x(\tau),x(\tau')) &=& - {1\/ 8 \pi^2} \left( {1\/(\tau-\tau'
	+ i \epsilon)^2} + {1\/(\tau-\tau' - i \epsilon)^2} \right),
	\label{eq34}\\
\chi^F(x(\tau),x(\tau')) &=& - {i \/ 4\pi (\tau-\tau')} \,\delta
(\tau-\tau').
	\label{eq35}\eea
The contributions (\ref{eq25}) and (\ref{eq26}) to spontaneous
emission
can now be evaluated using the statistical functions of the atom,
which are
given by (\ref{eq29}) and (\ref{eq30}). With a substitution $u=\tau -
\tau'$
we get:
\bea \left\langle {d H_A (\tau) \/ d\tau} \right\rangle_{vf} &=&
	{\mu^2 \/ 8 \pi^2} \sum_b \omega_{ab} |\langle a | R_2^f (0)
	|b\rangle |^2 \int_{-\infty}^{+\infty} du\, \left( {1\/
	(u+ i \epsilon)^2}+{1\/ (u-i \epsilon)^2} \right)
e^{i\omega_{ab}u},
	\label{eq36}\\
\left\langle {d H_A (\tau) \/ d\tau} \right\rangle_{rr} &=&
	-i{\mu^2 \/ 4 \pi} \sum_b \omega_{ab} |\langle a | R_2^f (0)
	|b\rangle |^2 \int_{-\infty}^{+\infty} du\, \delta' (u)
	e^{i \omega_{ab} u}, \label{eq37}\eea
where we have extended the range of integration to infinity for
sufficiently
long times $\tau-\tau_0$. We have also used the identity $\delta(u) =
- u \delta'(u)$.

After the evaluation of the integrals we obtain for the contribution
of the
vacuum fluctuations to the rate of change of atomic excitation energy
\beq \left\langle {d H_A (\tau) \/ d\tau} \right\rangle_{vf} =
	-{\mu^2 \/2 \pi} \left( \sum_{\omega_a > \omega_b} {1\/2}
\omega_{ab}^2
	|\langle a | R_2^f (0) | b \rangle |^2
	- \sum_{\omega_a < \omega_b} {1\/2} \omega_{ab}^2 |\langle a
| R_2^f (0)
	| b \rangle |^2\right). \label{eq38}\eeq
This result possesses an interesting interpretation
\cite{Dalibard82}.
Consider first the case when the atom is initially in the excited
state
($|a \rangle = |+ \rangle$).
Then only the first term $(\omega_a > \omega_b)$ contributes. The
vacuum
fluctuations lead to a de-excitation of the atom in the excited
state:
$\left\langle {d H_A (\tau) \/ d\tau} \right\rangle_{vf} < 0$. This
conforms
with the old heuristic picture of spontaneous emission as stimulated
emission
induced by vacuum fluctuations \cite{Weisskopf35}. If, on the other
hand, the
atom is initially in the ground state ($|a \rangle= |- \rangle$),
there is
only a contribution from the second term $(\omega_a<\omega_b)$. We
see that
vacuum fluctuations tend to excite an atom in the ground state:
$\left\langle {d H_A (\tau) \/ d\tau} \right\rangle_{vf}>0$. Note
that if
only the effects of vacuum fluctuations are taken into account, both
spontaneous excitation and de-excitation occur with equal frequency.
Although spontaneous excitation does not occur for inertial atoms,
this result should not come as a surprise if we take the heuristic
picture
seriously: since stimulated excitation and de-excitation have equal
Einstein B coefficients, vacuum fluctuations should stimulate atomic
excitation as well as de-excitation \cite{Milonni84}.

The contribution of radiation reaction to the change in the atom's
energy
becomes
\beq  \left\langle {d H_A (\tau) \/ d\tau} \right\rangle_{rr} =
	-{\mu^2 \/2 \pi} \left( \sum_{\omega_a > \omega_b} {1\/2}
\omega_{ab}^2
	|\langle a | R_2^f (0) | b \rangle |^2
	+ \sum_{\omega_a < \omega_b} {1\/2} \omega_{ab}^2 |\langle a
| R_2^f (0)
	| b \rangle |^2 \right). \label{eq39}\eeq
The effect of radiation reaction always leads to a loss of atomic
energy,
$\left\langle {d H_A (\tau) \/ d\tau} \right\rangle_{rr}<0$,
independent of whether the atom is initially in the ground or excited
state.
This can be compared with radiation reaction in the classical theory,
which
has the same effect and results in the instability of classical
atoms.

The total rate of change of the atomic excitation energy is obtained
by adding the contributions of vacuum fluctuations and radiation
reaction:
\beq \left\langle {d H_A \/ d\tau} \right\rangle_{tot} =
	\left\langle {d H_A  \/ d\tau} \right\rangle_{vf}
	+ \left\langle {d H_A \/ d\tau} \right\rangle_{rr}.
\label{eq40}\eeq
We observe that the effects of both contributions for an atom in the
ground state $(\omega_a < \omega_b)$ have equal magnitudes but
opposite
sign so that they exactly cancel:
\beq \left\langle {d H_A \/ d\tau} \right\rangle_{tot}   =
	 -{\mu^2 \/ 2\pi} \sum_{\omega_a > \omega_b} \omega_{ab}^2
	|\langle a | R_2^f (0) | b \rangle |^2 . \label{eq41}\eeq
In the ground state of an atom there is a balance between vacuum
fluctuations
and radiation reaction which establishes that no spontaneous
excitation to
higher levels is possible for inertial atoms
\cite{Dalibard82,Fain69},
only transitions to lower-lying levels (spontaneous emission) occur.
This balance also ensures the stability of the ground state: An
inertial
atom in its ground state does not radiate. Eq. (\ref{eq41}), shows,
on the
other hand, that for an atom in the excited state, the effects of
vacuum
fluctuations and radiation reaction add with equal contributions to
the
familiar phenomenon of spontaneous emission.

To calculate the Einstein A coefficient for the spontaneous emission
of
inertially moving atoms, we simplify Eq. (\ref{eq41}) by noting that
\beq \sum_{\omega_a < \omega_b} \omega_{ab}^2 |\langle a | R_2^f (0)
	|b\rangle |^2 \pm \sum_{\omega_a > \omega_b} \omega_{ab}^2
	|\langle a | R_2^f (0) | b \rangle |^2 =
	\cases{ {1\/4} \omega_0^2 \cr -{1\/2} \omega_0^2 \langle a|
R_3^f (0)
	|a \rangle. \cr} \label{eq42}\eeq
In order $\mu^2$, we can replace $\omega_0 \langle R_3^f \rangle$ by
$\langle H_A \rangle$ and obtain a differential equation for the
atomic
excitation energy:
\beq \left\langle {d H_A (\tau) \/ d\tau} \right\rangle_{tot} =
	- {\mu^2 \/8 \pi} \omega_0 \left( {1\/2} \omega_0 + \langle
H_A (\tau)
	\rangle\right). \label{eq43}\eeq
Note that no factors of $\gamma$ appear in (\ref{eq43}), because we
took
care to express all observables in the rest frame of the atom. The
solution of (\ref{eq43}) is
\beq \langle H_A (\tau) \rangle = - {1\/2} \omega_0 + \left( \langle
H_A (0)
	\rangle + {1\/2} \omega_0 \right) e^{-A \tau },
\label{eq44}\eeq
the familiar exponential decay to the atomic ground state $\langle
H_A
\rangle = -{1\/2}\omega_0$. The spontaneous emission rate is given by
the Einstein A coefficient of the scalar theory:
\beq A  = {\mu^2 \/ 8 \pi} \omega_0. \label{eq45}\eeq

How would the above results have been modified if the initial state
of the field were not the vacuum, but a bath of thermal radiation,
described
by a density matrix $\rho=\exp(-\beta H)$ with $\beta=1/kT$? The
formalism
can be easily generalized to that case. The vacuum expectation value
in (\ref{eq21}) and (\ref{eq22}) has simply to be replaced by a
reservoir average. The statistical functions of the field
(\ref{eq23})
and (\ref{eq24}) are then replaced by thermal Green functions:
\bea C^F_\beta (x(\tau),x(\tau')) = {1\/2} \hbox{Tr}\left( \rho
	\left\{ \phi(x(\tau)), \phi(x(\tau')) \right\}\right)
\label{eq45a}\\
 \chi^F_\beta (x(\tau),x(\tau')) = {1\/2} \hbox{Tr}\left( \rho \left[
 	\phi(x(\tau)), \phi(x(\tau')) \right]\right)
\label{eq45b}\eea
where the trace extends over the field degrees of freedom. The
contributions
of reservoir fluctuations and radiation reaction to the rate of
change of
the atom's excitation energy can be found according to (\ref{eq25})
and (\ref{eq26}) with the formulas (\ref{eq45a}) and (\ref{eq45b})
for the finite temperature statistical functions of the field.

The thermal Green functions (\ref{eq45a}) and (\ref{eq45b}) can be
expressed
in terms of the statistical functions for the vacuum. As mentioned
above, the
linear susceptibility of the field does not depend on the state of
the field
and is therefore equal to that of the vacuum:
\beq \chi^F_\beta (x(\tau),x(\tau')) = \chi^F (x(\tau),x(\tau')).
\label{eq45c}
\eeq

Hence, we have to deal only with the symmetric correlation function
$C^F_\beta$ of the field. It can be generally shown that $C^F_\beta$
is
connected to the vacuum symmetric correlation function by
(cf. \cite{Birrell82}, Eq. (2.111)):
\beq C^F_\beta (t(\tau),\x(\tau),t(\tau'),\x(\tau')) =
	\sum_{k=-\infty}^{+\infty} C^F_\beta
	(t(\tau)+ik\beta,\x(\tau),t(\tau'),\x(\tau')).
\label{eq45d}\eeq
This function appears in the expression (\ref{eq25}) for the
contribution
of reservoir fluctuations to the change of the atomic energy. The
terms
$k \neq 0$ in (\ref{eq45d}) represent the influence of the thermal
heat bath.

Let us consider an atom at rest in a thermal bath of radiation. As we
have
seen, the contribution of radiation reaction remains unchanged and is
given by (\ref{eq39}). To determine the contribution of reservoir
fluctuations, we have to calculate the finite temperature symmetric
correlation function. If we evaluate (\ref{eq45d}) for an atom at
rest in the
thermal bath (trajectory (\ref{eq33}) with $\vec v=0$, $\gamma=1$),
we obtain
\beq C^F_\beta (x(\tau),x(\tau')) = -{1\/8\pi^2}
\sum_{k=-\infty}^{+\infty}
	\left( {1\/(\tau-\tau'+ik\beta+i\epsilon)^2} +
	{1\/(\tau-\tau'+ik\beta-i\epsilon)^2}\right).
\label{eq45e}\eeq
We will not discuss here the evaluation of (\ref{eq25}) with the
thermal Green function (\ref{eq45e}). This will be postponed to
below, where
a comparison with the case of a uniformly accelerated atom can be
made.

\section{Uniformly accelerated atom}

Let us now generalize the preceding discussion to the case of a
uniformly
accelerating atom. We go back to the results of Sec. \ref{sec:rate}
and
specify the atom trajectory by
\beq t(\tau)= {1\/a} \sinh a \tau, \qquad z(\tau) = {1\/a} \cosh a
\tau,
\qquad x(\tau)=y(\tau) = 0, \label{eq46}\eeq
where $a$ is the proper acceleration. The statistical functions of
the field
for the trajectory (\ref{eq46}) can be evaluated from their general
forms (\ref{eq31}) and (\ref{eq32}). After some algebra, we obtain
\bea  C^F(x(\tau), x(\tau')) &=& -{a^2 \/32\pi^2} \left(
{1\/\sinh^2[{a\/2}
	(\tau-\tau') +i a \epsilon]}+ {1\/\sinh^2[{a\/2} (\tau-\tau')
-i a
	\epsilon]} \right) \nn\\
	&=& - {1\/8 \pi^2} \sum_{k=-\infty}^\infty \left(
{1\/(\tau-\tau' +
	{2 \pi i\/a} k+2 i \epsilon )^2}+{1\/(\tau-\tau' +
	{2 \pi i\/a} k - 2 i \epsilon )^2} \right) \label{eq47}\\
\chi^F(x(\tau),x(\tau'))  &=& -{i \/8\pi}{a\/\sinh{a\/2}(\tau-\tau')}
\,
	\delta (\tau-\tau'). \label{eq48}\eea
To obtain the second line, Eq. (4.3.92) from Ref. \cite{Abramowitz72}
has
been used. The stationarity of the motion of the atom is reflected by
the
fact that only the time difference $\tau$-$\tau'$ appears.
Comparison of (\ref{eq47}) with (\ref{eq45e}) and of (\ref{eq48})
with
(\ref{eq45c}) and (\ref{eq35}) shows complete agreement of the
statistical
functions $C^F$ and $\chi^F$ of the field for a trajectory with
$a=\hbox{const}$ through the vacuum on one hand and an inertial
trajectory
which is at rest with respect to a thermal bath with temperature
$T=a/2\pi$ on the other. This well-known fact has consequences for
the
separate discussion
of the influence of vacuum fluctuations and of radiation reaction on
the
spontaneous excitation and de-excitation of an accelerated atom.

The contribution of the vacuum fluctuations to the rate of change of
the
atomic Hamiltonian becomes with (\ref{eq25})
\bea \left\langle {d H_A (\tau) \/ d\tau}\right\rangle_{vf} &=&
	{\mu^2 \/ 8 \pi} \sum_b \omega_{ab}  |\langle a | R_2^f
(0)|b\rangle |^2
	\nn\\
	 &&\times \sum_{k=-\infty}^\infty \int_{-\infty}^\infty du \,
\left[
	{e^{i \omega_{ab} u} \/ (u + {2\pi i\/a} k + 2 \pi i \epsilon
)^2 }
	+{e^{i \omega_{ab} u} \/ (u + {2\pi i\/a} k - 2 \pi i
\epsilon )^2 }\right].
	\label{eq49}\eea
Again, we have extended the range of integration to infinity for
sufficiently
long times. It is interesting to compare (\ref{eq49}) with the
corresponding
expression (\ref{eq36}) for the inertial atom. There, only the term
$k=0$
was present. The remaining terms describe the modification due to the
acceleration. The integrals can be calculated using the residue
theorem,
leading to a geometric series for the $k$ summation. The resulting
expression for the rate of change of atomic excitation energy caused
by
vacuum fluctuations is
\bea \left\langle {d H_A (\tau) \/ d\tau} \right\rangle_{vf} &=&
	-{\mu^2\/ 2\pi} \Biggl[ \sum_{\omega_a > \omega_b}
\omega_{ab}^2
	|\langle a | R_2^f (0) | b \rangle |^2
	\left( {1\/2} + {1\/ e^{{2 \pi\/a} \omega_{ab}} -1} \right)
\nn\\
	&& -\sum_{\omega_a < \omega_b} \omega_{ab}^2 |\langle a |
R_2^f (0) |
	b \rangle |^2 \left( {1\/2} + {1\/ e^{{2 \pi\/a}
|\omega_{ab}|} -1}
	\right) \Biggr]. \label{eq50}\eea
We note the appearance of the thermal terms in addition to the
inertial
vacuum fluctuation terms $1\/2$. As for an inertial atom, vacuum
fluctuations
tend to excite an accelerated atom in the ground state and de-excite
it
in the excited state. Both processes are supported with equal
magnitude,
and are enhanced by the thermal terms compared to the inertial case.

Turning to the contribution of radiation reaction, we find with
(\ref{eq26})
\beq  \left\langle {d H_A (\tau) \/ d\tau} \right\rangle_{rr} =
	i {\mu^2 \/ 8 \pi} \sum_b \omega_{ab} |\langle a | R_2^f (0)
| b\rangle
	|^2 \int_{-\infty}^\infty du {a\/ \sinh {a\/2} u} \,\delta
(u) \,
	e^{i \omega_{ab} u}. \label{eq51}\eeq
After the evaluation of the integral, this becomes
\beq \left\langle {d H_A (\tau) \/ d\tau} \right\rangle_{rr}=
	- {\mu^2 \/ 2 \pi} \left(\sum_{\omega_a > \omega_b} {1\/2}
\omega_{ab}^2
	|\langle a | R_2^f (0) | b \rangle |^2 + \sum_{\omega_a <
\omega_b} {1\/2}
	\omega_{ab}^2 |\langle a | R_2^f (0) | b \rangle |^2 \right).
	\label{eq52}\eeq
It is remarkable that this is the same expression as (\ref{eq39}) for
an
inertial atom, leading again always to a loss of energy. A similar
situation is known in classical electrodynamics: a uniformly
accelerated
charge on the trajectory (\ref{eq46}) is not subject to a radiation
reaction force, although radiation is emitted \cite{Ginzburg79}.
Thus the fact that the contribution of radiation reaction is not
changed
in the uniformly accelerated case is probably a property of the
particular trajectory (\ref{eq46}).

Finally, we add the contributions of vacuum fluctuations (\ref{eq50})
and
radiation reaction (\ref{eq52}) to obtain the total rate of change of
the atomic excitation energy:
\bea \left\langle {d H_A \/ d\tau} \right\rangle_{tot} &=& {\mu^2 \/
2\pi}
	\Biggl[-  \sum_{\omega_a > \omega_b} \omega_{ab}^2 |\langle a
	| R_2^f (0) | b \rangle |^2 \left( 1 +{1\/ e^{{2\pi \/a}
\omega_{ab} }-1}
	\right) \nn\\
	&\qquad&\qquad + \sum_{\omega_a < \omega_b} \omega_{ab}^2
|\langle a
	| R_2^f (0)|b \rangle |^2 {1\/ e^{{2\pi \/a} |\omega_{ab}|
}-1}
	\Biggr] . \label{eq53}\eea
For an atom in the excited state, only the term $\omega_a>\omega_b$
contributes. It describes the {\it spontaneous emission} of an
accelerated
atom. Compared to an inertial atom, it is modified by the appearance
of the
thermal term. If, however, the atom is in the ground state, there is
a
nonzero contribution from the term $\omega_a<\omega_b$. For an atom
in its
ground state in uniformly accelerated motion through the Minkowski
vacuum
there is
no perfect balance between vacuum fluctuations and radiation
reaction.
Accordingly, transitions to the excited state
($\left\langle {d H_A (\tau) \/ d\tau} \right\rangle_{tot}>0$) become
possible even in the vacuum. This {\it spontaneous excitation} is the
Unruh effect \cite{Unruh76} which has now been traced back to the
interplay between
the two underlying physical effects. The often stated conjecture that
the
Unruh effect goes back to the vacuum fluctuations has been made
precise
in this sense.

Note that for an atom in the ground state, the atomic energy can only
increase. A loss of energy from the ground state, which would have
fatal
consequences for the stability of the atom, is not possible according
to (\ref{eq53}).

\section{Evolution of the atomic population, Einstein coefficients}

Analogous to the procedure in Sec. \ref{sec:inertial}, we can
simplify
Eq. (\ref{eq53}) to obtain a differential equation for $\langle H_A
\rangle$:
\beq \left\langle {d H_A (\tau) \/ d\tau} \right\rangle = -{\mu^2 \/
4\pi}
	\omega_0 \left( {1\/4} \omega_0 + \left( {1\/2} + {1\/
e^{{2\pi \/a}
	\omega_0} -1} \right)\langle H_A (\tau) \rangle \right).
\label{eq54}\eeq
Its solution gives the time evolution of the mean atomic excitation
energy:
\bea \langle H_A (\tau) \rangle &=& - {1\/2} \omega_0 + { \omega_0 \/
	e^{{2\pi \/a} \omega_0} +1} \nn\\
	&& + \left( \langle H_A (0) \rangle + {1\/2} \omega_0 -  {
\omega_0 \/
	e^{{2\pi \/a} \omega_0} +1}\right) \, \exp\left[ -{\mu^2\/4
\pi} \omega_0
	\left( {1\/2} + { 1 \/ e^{{2\pi \/a} \omega_0} -1} \right)
\tau \right].
	\label{eq55}\eea
We see that the atom evolves with a modified decay parameter towards
the
equilibrium value
\beq \langle H_A  \rangle  = - {1\/2} \omega_0 + { \omega_0 \/
	e^{{2\pi \/a} \omega_0} +1},\label{eq56}\eeq
representing a thermal excitation with temperature $T=a/2\pi$ above
the
ground state. It is remarkable that the atom obeys Fermi-Dirac
statistics
in thermal equilibrium. This is due to the fermionic nature of a
two-level system (for example, the atomic raising and lowering
operators
obey the anticommutation relation $\{R_+,R_-\} =1$).

The identification of the Einstein coefficient is more complicated
than
in the inertial case, since we have two competing processes now, that
occur both spontaneously. There are two Einstein coefficients
$A_\downarrow$ and $A_\uparrow$ which describe the transition rates
corresponding to these processes.

Einstein coefficients are defined with respect to rate equations for
the
atomic populations. Consider therefore an ensemble of $N$ atoms. Let
$N_1$ denote the number of atoms in the ground state, $N_2$ the
number
in the excited state, with $N=N_1+N_2$.. The rate equations
describing
the two spontaneous processes above are
\beq {d N_2 \/ d \tau} = - {d N_1 \/ d \tau} = A_{\uparrow} N_1 -
	A_{\downarrow} N_2 \label{eq57}\eeq
with
\beq \langle H_A \rangle = {1\/N} \left( - {1\/2} \omega_0 N_1 +
{1\/2}
	\omega_0 N_2 \right). \label{eq58}\eeq
The solution of the rate equations,
\beq  \langle H_A (\tau) \rangle = - {1\/2} \omega_0 + \omega_0
	{ A_{\uparrow} \/ A_{\uparrow} + A_{\downarrow}} + \left(
\langle H_A (0)
	 \rangle + {1\/2} \omega_0 - { A_{\uparrow} \/ A_{\uparrow}
	 +A_{\downarrow}}\omega_0 \right) e^{- ( A_{\uparrow} +
A_{\downarrow} )
	 \tau }, \label{eq59}\eeq
shows also an equilibrium state different from the ground state and a
modified decay constant. Comparing (\ref{eq55}) and (\ref{eq59}),
we can identify the Einstein coefficients $A_\downarrow$ and
$A_\uparrow$
for an accelerated atom:
\beq  A_{\downarrow} = {\mu^2 \/ 8 \pi} \omega_0 \left( 1 + {1\/
e^{{2 \pi
	\/a} \omega_0} -1} \right), \qquad A_{\uparrow} = {\mu^2 \/ 8
\pi}
	\omega_0 {1\/ e^{{2 \pi \/a} \omega_0}-1}. \label{eq60}\eeq
The coefficient $A_\downarrow$ for spontaneous emission from an
accelerated atom can be compared to its inertial value (\ref{eq45}).
We see that the rate of spontaneous emission is enhanced by the
thermal
contribution. The transition rate $A_\uparrow$ for the spontaneous
excitation,
on the other hand, is given by (\ref{eq45}) weighted with the thermal
factor.
It vanishes as $a\to 0$.

\section{Conclusions}

We have studied atoms in arbitrary stationary motion
through the vacuum of a photon field. For an atom observable $G$, the
total
rate $dG\/d \tau$ with regard to the proper time $\tau$ of the atom
can be
split into two distinct rates going back to vacuum fluctuations
acting on
the atom and to radiation reaction. This splitting is unique if one
demands that the two processes should separately have a physical
meaning.
Whether spontaneous processes may occur for the different states of
the moving atoms, depends on on the balance between the two physical
processes.
The corresponding symmetric correlation function and linear
susceptibility of
the field are simply obtained by calculating the respective special
relativistic Green functions with respect to the proper time along
the
atom's trajectory.

We computed the two contributions to the variation of the mean energy
of a state of a two-level atom for uniform motion and for constant
acceleration. For the ground state of an atom in uniform motion, the
influence of vacuum fluctuations is exactly canceled by the radiation
reaction. If an atom moves with constant acceleration, this perfect
balance
is disturbed. The radiation reaction remains unchanged, but the
vacuum
fluctuations are modified by an additional term with thermal spectrum
of temperature $T=\hbar a/2\pi c k=4\cdot 10^{-20}K (a/9.81
{{m}\/{{s}^2}})$. This has two consequences: spontaneous excitation
from the ground state becomes possible (Unruh effect) and the
Einstein
coefficient for spontaneous emission from the upper state is changed.
These effects are now traced back quantitatively to the two
underlying physical processes.

The calculations presented above can be easily transcribed to other
physical
situations. The change to another atom trajectory in Minkowski space
is obvious. In addition, continuous efforts over the last two decades
have provided us with a `zoo' of worked out vacuum expectation values
of products of free field operators (Green functions) for different
fields in various curved spacetimes and in situations with
boundaries.
We mention black holes \cite{Candelas80}, moving mirrors
\cite{Birrell82},
cosmic strings \cite{Davies88}, and Robertson-Walker universes
\cite{Birrell82}. In all these cases, the two distinct effects
underlying
spontaneous processes of atoms on certain trajectories can be worked
out
following closely the scheme presented above.

\vskip 0.5cm

\noindent
{\bf Acknowledgments}\\
This work was supported by the Studienstiftung des deutschen Volkes.


\end{document}